\documentstyle[epsfig,12pt,a4p]{article}
\newcommand{\be}{\begin{equation}} \newcommand{\ee}{\end {equation}}

\newcommand{\pom}{ $I\hspace{-1.6mm}P$}
\begin{document}
\bibliographystyle{unsrt}
\def\question#1{{{\marginpar{\small \sc #1}}}}

%\rightline{hep-ph/0000000}
% \rightline{RAL-00-001}
\rightline{5 June 2000}
\baselineskip=18pt
\vskip 0.7in
\begin{center}
{{\bf \LARGE Glueballs: A central mystery}}\\
\vspace*{0.9in}
{\large Frank E. Close}\footnote{\tt{e-mail: F.E.Close@rl.ac.uk}} \\
\vspace{.1in}
{\it CERN, Geneva, Switzerland}\\
{\it and}\\
{\it Rutherford Appleton Laboratory}\\
{\it Chilton, Didcot, OX11 0QX, England}\\
%{\large Andrew Kirk}\footnote{\tt{e-mail: ak@hep.ph.bham.ac.uk}} \\
%{\it School of Physics and Space Research}\\
%{\it Birmingham University}\\
%\vspace{0.1in}
%{\large Gerhard Schuler}\footnote{\tt{e-mail: Gerhard.Schuler@cern.ch}} \\
%{\it CERN, Geneva, Switzerland}\\
%\vspace*{0.1in}
\end{center}
%\maketitle
%\vspace{-0.25cm}
\begin{abstract}
Glueball candidates and $q\bar{q}$ mesons have been
found to be produced with
 different momentum and angular dependences
in the central region
of $pp$ collisions. This talk illustrates
this phenomenon and
 explains the $\phi$ and $t$ dependences of
mesons with $J^{PC} = 0^{\pm +},1^{++},2^{\pm +}$.
For production of $0^{++}$ and $2^{++}$ mesons the analysis reveals a
systematic behaviour in the data that appears to distinguish between
$q\bar{q}$ and non-$q\bar{q}$ or glueball candidates. An explanation is given
for the absence of $0^{-+}$ glueball candidates in central production at
present energies and
the opportunity for their discovery at RHIC is noted.
\end{abstract}
\newpage
\setcounter{footnote}{0}

The idea that glueball production might be favoured in the central region
of $pp \rightarrow pMp$ by the fusion of two
Pomerons (\pom ) is over twenty years old~\cite{robson,fcrpp}. The
fact that known $q\bar{q}$ states also are seen in this process frustrated
initial hopes that such experiments would prove to be a clean glueball
source. However, in \cite{ck97} we noted a kinematic effect whereby
known $q\bar{q}$ states could be suppressed leaving potential glueball
candidates more prominent.

Its essence was that the pattern of resonances produced in the central
region of double tagged $pp \rightarrow pMp$ depends on the vector
$ difference$ of the transverse momentum recoil of the final state
protons (even at fixed four momentum transfers). When this quantity
($dP_T \equiv |\vec{k_{T1}} - \vec{k_{T2}}|$) is large, ($\geq O(\Lambda
_{QCD})$), $q\bar{q}$ states are prominent whereas at small
$dP_T$ all well established $q\bar{q}$ are observed to be suppressed
while the surviving resonances include the enigmatic $f_0(1500),
f_0(1710)$ and $f_0(980)$.

The data are consistent with the hypothesis that as $dP_T \rightarrow 0$
all bound states with internal $L > 0$ (e.g. $^3P_{0,2}$ $q\bar{q}$)
are suppressed while S-waves survive (e.g. $0^{++}$ or $2^{++}$ glueball
made of vector gluons and the $f_0(980)$ as any of glueball,
or S-wave $qq\bar{qq}$ or $K \bar{K}$ state). Models are
needed to see if such a pattern is natural.
Following this discovery there has been an intensive experimental
programme in the last two years by the WA102 collaboration at CERN, which
has produced a large and detailed set of data on both the
$dP_T$
{}~\cite{ck97} and the azimuthal angle,
$\phi$,
dependence of meson
production (where
$\phi$ is the angle between the transverse momentum
vectors, $p_T$, of the two outgoing protons).
\par
The azimuthal dependences
as a function of
$J^{PC}$ and the momentum transferred at the proton vertices, $t$,
are very striking.
As seen in refs.~\cite{WAphi,WA2-+}, and
later in this paper, the $\phi$ distributions for mesons with
$J^{PC}$~=~$0^{-+}$ maximise around $90^{o}$, $1^{++}$ at $180^o$ and $2^{-+}$
at $0^o$.
Recently, the WA102 collaboration has confirmed that this is not
simply a J-dependent effect~\cite{pi4papr} since
$0^{++}$ production peaks at $0^o$ for some states whereas others are more
evenly spread~\cite{WA0++}; $2^{++}$ established $q\bar{q}$ states peak at
$180^o$ whereas the $f_2(1950)$, whose mass may be consistent with the tensor
glueball predicted in lattice QCD, peaks at $0^o$~\cite{pi4papr}.

In this talk I show how these
phenomena arise and in turn expose the extent to which they could be
driven, at least in part, by the internal structure of the meson in
question and thereby be exploited as a glueball/$q\bar{q}$
filter~\cite{ck97}.
We find that the $\phi$ dependences of $0^{-+}$
and
$1^{++}$ follow on rather general grounds if a single trajectory
dominates the production mechanism. Having thus established the
ability to describe the phenomena quantitatively in these cases, we
predict the behaviour for
$2^{-+}$ production and then confront the $0^{++}$ and  $2^{++}$
$glueball/q\bar{q}$
sector.

To orient ourselves, think of $e^+ e^- \rightarrow e^+ e^- M$ where
the essential production dynamics is through $\gamma \gamma \rightarrow M$
 fusion. The photon can be polarised either $T$ ($\lambda = \pm 1$) or
$L$ ($\lambda = 0$). For $J^{PC} = 0^ {++}$ the resulting structure is
$(R-cos(\phi))^2$ where $R$ is equal to the ratio of
the longitudinal and transverse production amplitudes for
$\gamma \gamma \rightarrow M$ and depends on the dynamical
structure of the meson, $M$. By contrast,
parity forbids the production of $0^{-+}$ by the fusion of two scalars
and also by the longitudinal ($``L"$) components of two vectors.
Transverse ($``T"$)  components are allowed and so a single
amplitude drives the $\gamma \gamma$ fusion in production of the
$0^{-+}$ states. The resulting distribution is predicted to behave like
$sin^2 (\phi)$.

In ref. \cite{cs1} it was noted that these distributions are very similar
to what is found experimentally
in  $pp \rightarrow pMp$ and so a CVC model for the Pomeron
was used to confront the data for a range of mesons, $M$.
The results were very similar, but not identical, to the data, e.g.
the $^3P_2$ $q \bar{q}$ states are produced dominantly with
$\lambda = 0$ in \pom \pom fusion instead of $\lambda = \pm 2$
in $\gamma \gamma$ fusion. With hindsight the reason is obvious:
\pom is not a $conserved$ vector current and, in effect, has an intrinsic
(and important) scalar component. One effect is that whereas amplitudes for
longitudinal $\gamma$ emission are suppressed as $t \to 0$, the case for
the analogous \pom $grows$. Suddenly everything began to fit, as
summarised in ref.\cite{cs2} and in the experimental paper
`Experimental evidence for a vector-like behaviour of Pomeron
      exchange'  \cite{WAphi}.
I will now show how the data can be quantitatively
described in this simple picture and how characteristic
features that may discriminate glueball from $q \bar{q}$
states may ensue.

\noindent $J^{PC}=0^{-+}$

The detailed calculations are described in~\cite{cs1,cs2}.
Here I shall concentrate on the $\eta^\prime$ meson
whose production has been found to be consistent with double
pomeron exchange~\cite{WAphi}. The resulting
behaviour of the cross section may be summarised as follows:

\[
\frac{d\sigma}{dt_1 dt_2 d \phi^\prime}
\sim t_1 t_2 {G^p_E}^{2} (t_1)
{G^{p}_{E}}^2 (t_2) \sin^2(\phi^\prime) F^2(t_1, t_2, M^2)
\]
% \noindent where $F(t_1, t_2, M^2)$ is the \pom-\pom-$\eta^\prime$
\noindent where $\phi^\prime$ is the angle between the two $pp$ scattering
planes in the \pom-\pom\thinspace centre of mass
and $F(t_1, t_2, M^2)$ is the \pom-\pom-$\eta^\prime$
form factor. We
temporarily set this equal to unity; $pp$ elastic scattering data
and/or a Donnachie Landshoff type form factor~\cite{dl} can be used as
model of the proton-\pom\thinspace form factor ($G_E^p(t)$).

% An intuitive explanation for this distortion may be seen to come from
% the flux factors $(x_1x_2s)^{-1} \rightarrow M_T^2$ where $M_T^2 = M^2
% + (1-x_1)|t_1| + (1-x_2)|t_2| + 2
% \sqrt{((1-x_1)(1-x_2)t_1t_2}cos(\phi)$ with $cos(\phi) \sim q_{1T}
% \cdot q_{2T}$. This leads to a $\phi$ dependent and $t$ dependent
% distortion. Hence the $\phi^\prime$ distribution
% for ``real" kinematics in the current - current c.m. is a convolution
% of the original ``symmetric" $\sin^2(\phi^\prime)$~\cite{cs1}  with the
% above ``hidden" $\phi$ dependence.

The WA102 collaboration measures the azimuthal angle ($\phi$)
in the $pp$ c.m. frame and so we
transform the $\phi^\prime$ from the current c.m. frame to $\phi$ for the
$pp$ c.m. frame. To generalise to real kinematics, we use
a Monte Carlo simulation based on Galuga
\cite{galuga} modified for $pp$ interactions and incorporating
the \pom-proton form factor from ref.~\cite{dl}.

In order to fit the data we found
that the \pom-\pom-meson form factor $ F(t_1, t_2, M^2) $
% that the Pom-Pom-meson form factor $ F(t_1, t_2, M^2) $
has to differ from unity.
If we
parametrise $ F^2(t_1, t_2, M^2) $
as $exp^{- b_T(t_1+t_2)}$ then we need $b_T$~=~0.5~$GeV^{-2}$
in order to describe the $t$
dependence.
Fig. (1a and 1b) compare the final theoretical form for the $\phi$
distribution and the $t$ dependence with the data for
the $\eta^\prime$; (the distributions are well described also for
the $\eta$ but it
has not yet been established that \pom-\pom \thinspace alone dominates the
%has not yet been established that Pom-Pom alone dominates the
production of this meson).

The $t_1t_2$ factors in the cross section arise from the $TT$ nature of the
amplitude
and will be general for the production of any $0^{-+}$ meson. Hence for
$0^{-+}$ states with $M >> 1 $GeV, as expected for the lattice glueball or
radial
excitations of $q \bar{q}$, this dynamical $t_1t_2$ factor will suppress
the region where kinematics would favour the production.  It would be
interesting if
glueball production dynamics involved a
singular $(t_1t_2)^{-1}$ that compensated for the transverse \pom factor, as
in this case the cross section would stand out. However, we have no reason
to expect such a fortunate accident. Hence observation of high mass
$0^{-+}$ states is expected only to be favourable at extreme energies, such as
at RHIC
or LHC.

\noindent $J^{PC} = 1^{++}$

In refs.~\cite{cs1,cs2,fc1+} Close and Schuler have predicted that axial mesons
are produced polarised, dominantly in helicity one; this is verified by
data
\cite{WA1+pol}.  The cross section is predicted to have the form

\[
\frac{d\sigma}{dt_1dt_2d\phi^\prime} \sim t_1t_2 [ \{A(t_1^T,t_2^L) -
A(t_2^T,t_1^L)\}^2 + 4 A(t_1^T,t_2^L)A(t_1^L,t_2^T)\sin^2(\phi^\prime/2) ]
\]

% where $A(t_i,t_j)$ are the Pom-Pom-$f_1$ form factors.
\noindent where $A(t_i,t_j)$ are the \pom-\pom-$f_1$ form factors.
In the models of refs.~\cite{cs2,adln} the longitudinal Pomeron
amplitudes carry a factor of $1/\sqrt{t}$ arising from the fact that, in
the absence of any current conservation for the Pomeron, a longitudinal
vector polarisation is not compensated. Thus we make this factor explicit
and write $A(t_i,t_j^L) = \frac{\mu}{\sqrt{t_j}} a(t_i,t_j)$; hence

\[
\frac{d\sigma}{dt_1dt_2d\phi^\prime} \sim
[\{\sqrt{t_1} - \sqrt{t_2}\frac{a(t_1^T,t_2^L)}{a(t_1^L,t_2^T)}\}^2 + 4
\sqrt{t_1t_2} \frac{a(t_1^T,t_2^L)}{a(t_1^L,t_2^T)}
\sin^2(\phi^\prime/2)]a^2(t_1^L,t_2^T)
\]

In the particular case where the ratio of form factors is unity, this
recovers the form used in ref.~\cite{cs2}

\[
\frac{d\sigma}{dt_1dt_2d\phi^\prime} \sim
[(\sqrt{t_1} - \sqrt{t_2})^2 + 4 \sqrt{t_1t_2}
\sin^2(\phi^\prime/2)]a^2(t_1,t_2)
\]

\noindent which implies a dominant $\sin^2(\phi/2)$ behaviour that tends to
isotropy when suitable cuts on $t_i$ are made. This is qualitatively
realised (figs.~1e and f of ref.~\cite{WAphi}).

We have
parametrised
$a(t_i^T,t_j^L)$
as an exponential,
$exp^{-(b_Tt_i+b_Lt_j))}$ where $i,j=1,2$; $b_T$~=~0.5 $GeV^{-2}$ was
determined from the
$\eta^\prime$ data above; $b_L$ is
determined from
the overall $t$ dependence of the $1^{++}$ production and requires
$b_L$~=~3~$GeV^{-2}$.
Fig.~(2a and b) show the output of the model predictions from the
Galuga Monte Carlo superimposed
on the $\phi$ and $t$ distributions for the $f_1(1285)$ from
the WA102 experiment.

In addition we have a parameter free
prediction of
the variation of the $\phi$ distribution as a function of
$|t_1-t_2|$.
Fig.~(2c and d) show the output of the Galuga Monte Carlo superimposed
on the $\phi$ for the $f_1(1285)$ for $|t_1 - t_2|$~$\le$~0.2~$GeV^{-2}$
and $|t_1 - t_2|$~$\ge$~0.4~$GeV^{-2}$ respectively.
The agreement between the data and
our prediction is excellent.
Similar conclusions arise for the $f_1(1420)$.

\noindent $J^{PC}=2^{-+}$

The $J^{PC}$~=~$2^{-+}$ states, the $\eta_2(1645)$ and $\eta_2(1870)$,
are predicted to be produced
polarised. Helicity 2 is suppressed by Bose symmetry~\cite{cs1} and
has been found to be negligible experimentally~\cite{WA2-+}.
The structure of the cross section
is then predicted to be

\noindent (i) helicity zero:  as for the $0^{-+}$ case,
\[
\frac{d\sigma}{dt_1dt_2d\phi^\prime} \sim t_1t_2 \sin^2(\phi^\prime)
\]

\noindent (ii) helicity one:

\[
\frac{d\sigma}{dt_1dt_2d\phi^\prime} \sim
[\{\sqrt{t_1} - \sqrt{t_2}\frac{a(t_1^T,t_2^L)}{a(t_1^L,t_2^T)}\}^2 + 4
\sqrt{t_1t_2} \frac{a(t_1^T,t_2^L)}{a(t_1^L,t_2^T)}
\cos^2(\phi^\prime/2)]a^2(t_1^L,t_2^T)
\]
\noindent which is as the $1^{++}$ case except for the important and
significant change from $\sin^2(\phi^\prime/2)$ to $\cos^2(\phi^\prime/2)$.

 The
uncompensated factor of $t_1t_2$ in the helicity zero component
tends to suppress this kinematically under the conditions of the WA102
experiment. Indeed, WA102 find that helicity one alone is able to
describe their data~\cite{WA2-+}; this is in interesting contrast to
$\gamma
\gamma
\to \eta_2(Q\bar{Q})$ in the non-relativistic quark model where the
helicity-one amplitude would be predicted to vanish~\cite{fczpli}. We shall
concentrate on this helicity-one amplitude henceforth.

The results of the WA102 collaboration for the $\eta_2(1645)$~\cite{WA2-+}
are shown
in fig.~(3a and b).
The distribution peaks as $\phi \to 0$, in
marked contrast to the suppression in the $1^{++}$ case (fig.~2a).

Integrating our formula over $\phi$, with the same approximations as
previously, implies

\[
\frac{d\sigma}{dt_1dt_2} \sim (t_1 + t_2) ( exp^{-(b(t_1+t_2)})
\]

\noindent and, in turn, that

\begin{equation}
\frac{d\sigma}{dt} \sim (1 + bt) ( exp^{-bt})
\label{eq:a}
\end{equation}

\noindent This simple form compares remarkably well with WA102 who fit to
$\alpha e^{-b_1t} + \beta t e^{-b_2t}$; our prediction (eq.~\ref{eq:a}) implies
that $b_1 \equiv b_2$ and that $\beta/\alpha \equiv b $ and WA102
find for the $\eta_2(1645)$)~\cite{WA2-+}
$b_1 = 6.4 \pm 2.0  ; b_2 = 7.3 \pm 1.3 $ and $\beta =2.6 \pm 0.9$,
$\alpha = 0.4 \pm 0.1$

Performing the
detailed comparison of model and data via Galuga, as in
the previous examples, leads to the results shown in fig.~(3a and b) for
the $\eta_2(1645)$; the $\eta_2(1870)$ results are qualitatively similar.
Bearing in mind that there are no free parameters, the
agreement is remarkable. Indeed, the successful
description of the $0^{-+}$, $1^{++}$ and $2^{-+}$ sectors, both
qualitatively and in detail, sets the scene for our
analysis of the $0^{++}$ and $2^{++}$ sectors where glueballs are
predicted to be present together with established $q\bar{q}$ states.
Any differences between data and
this model may then be
a signal for hadron structure, and potentially a filter for glue degrees of
freedom.

Before turning to the $0^{++},2^{++}$ channels with glueball interest,
 it is worth summarising exactly what we have assumed and what
we have described, parameter free.

For the production of $J^{PC}$~=~$0^{-+}$ mesons
we have predicted the $\phi$ dependence and
the vanishing cross section as $t \rightarrow 0$ absolutely and
have fitted the
$t$ slope in terms of one parameter,
$b_T$.
For the $J^{PC}$~=~$1^{++}$ mesons
we predict the general form for the $\phi$
distribution: it is in this channel for the first time
that the non-conserved nature of the \pom first manifests
itself. The polarisation of the $1^{++}$ is also natural.
By fitting the $t$ slope we obtain the parameter
$b_L$; this then gives a parameter free prediction for
the variation of the $\phi$ distribution as a function of $t$ which
agrees with the data.
With parameters now fixed, we obtain
absolute predictions for both the $t$ and $\phi$ dependences of the
$J^{PC}$~=~$2^{-+}$ mesons which are again in accord with the data
when helicity 1 dominance is imposed.

The message is that the production of the unnatural
spin-parity states, $0^{-+},1^{++},2^{-+}$, is driven by
the non-conserved vector nature of the exchanged \pom; it is
not immediately affected by the internal structure of the produced
meson. In particular, it is not sensitive to whether the
mesons are glueballs or $q \bar{q}$.

Now I shall look at the $0^{++}$ and $2^{++}$ sector where
glueballs are expected as well as $q\bar{q}$. Here we
shall find that the production topologies do depend on
the internal dynamics of the produced meson and as such may
enable a distinction between $q\bar{q}$ and exotic, glueball, states.

\newpage

\noindent $J^{PC}=0^{++}$ and $2^{++}$

In contrast to the $0^{-+}$ case, where parity forbade the LL amplitude,
in the $0^{++}$ case both $TT$ and $LL$ can occur. Hence there are two
independent form factors~\cite{cfl} $A_{TT}(t_1,t_2,M^2)$ and
$A_{LL}(t_1,t_2,M^2)$. For
$0^{++}$ and the helicity zero amplitude of $2^{++}$ (which
experimentally is found to dominate~\cite{WAhel0}) the angular dependence
of scalar meson production will be~\cite{cs2}

\begin{equation}
\frac{d\sigma}{dt_1 dt_2 d \phi^\prime}
\sim  {G^p_E}^{2} (t_1)
{G^{p}_{E}}^2 (t_2)[1 +
\frac{\sqrt{t_1t_2}}{\mu^2}\frac{a_T}{a_L}e^{(b_L-b_T)(t_1+t_2)/2 }
\cos(\phi^\prime)]^2 e^{-b_L(t_1+t_2) }
\label{eq:b}
\end{equation}

\noindent where we have written $a_L(t) = a_Le^{-(b_Lt/2)}$ and
$a_T(t) = a_Te^{-(b_Tt/2)}$ with $b_{L,T}$ fixed
to the values found earlier. The ratio
$a_T/a_L$ can be positive or negative, or in general even complex.

  Eq.(\ref{eq:b}) predicts that there should be significant changes
in the $\phi$ distributions as $t$ varies.
When
$\frac{\sqrt{t_1t_2}}{\mu^2}a_T/a_L \sim \pm1$, the $\phi$ distribution will be
$\sim \cos^4(\frac{\phi}{2})$ or $\sin^4(\frac{\phi}{2})$
depending on the sign.
Indeed data on the
enigmatic scalars $f_0(980)$ and $f_0(1500)$ show a
$\cos^4(\frac{\phi}{2})$ behaviour when
$\sqrt{t_1t_2} \leq 0.1$ GeV$^2$,
changing to $\sim \cos^2(\phi)$ when $\sqrt{t_1t_2} \geq
0.3$ GeV$^2$~\cite{WAphi}.

The overall $\phi$
dependences for the $f_0(1370)$, $f_0(1500)$, $f_2(1270)$ and
$f_2(1950)$ can be described by varying the quantity $\mu^2a_L/a_T$.
Results are shown in fig.~4.
It is clear that these $\phi$
dependences discriminate two classes of meson in the $0^{++}$ sector and
also in the $2^{++}$.
The $f_0(1370)$ can be described using $\mu^2a_L/a_T$~=~-0.5~$GeV^2$,
for the $f_0(1500)$ it is +0.7~$GeV^2$,
for the $f_2(1270)$ it is -0.4~$GeV^2$ and
for the $f_2(1950)$ it is +0.7~$GeV^2$.

It is interesting to note that we can fit these $\phi$ distributions
with one parameter and it is primarily the sign of this quantity that
drives the $\phi$ dependences.
Understanding the dynamical origin of this sign is now a central issue
in the quest to
distinguish $q \overline q$ states from
glueball or other exotic states.

\begin{center}
{\bf Acknowledgements}
\end{center}
\par
This is based on work performed in collaborations with
A.Kirk and G.Schuler and is supported, in part, by
%the European Community Human Mobility Program Eurodafne,
%contract NCT98-0169 and
the EU Fourth Framework Programme contract Eurodafne, FMRX-CT98-0169.

\clearpage
{ \large \bf Figures \rm}
\begin{figure}[h]
\caption{
%The predicted $\phi^\prime$ distributions for $J^{PC}$~=~$0^{-+}$ mesons
%a) naive distribution and b) taking into account the experimental kinematics.
a) The $\phi$
and b) the $|t|$ distributions for the $\eta^\prime$
for the data (dots) and the model predictions from the Monte Carlo (histogram).
}
\label{fi:1}
\end{figure}
\begin{figure}[h]
\caption{
a) The $\phi$
and b) the $|t|$
distributions for the $f_1(1285)$
for the data (dots) and the Monte Carlo (histogram).
c) and d) the $\phi$ distributions for $|t_1 - t_2|$~$\le$~0.2 and
$|t_1 - t_2|$~$\ge$~0.4~$GeV^{2}$ respectively.
}
\label{fi:2}
\end{figure}
\begin{figure}[h]
\caption{
a) The $\phi$
and b) the $|t|$
distributions for the $\eta_2(1645)$
for the data (dots) and the Monte Carlo (histogram).
}
\label{fi:3}
\end{figure}
\begin{figure}[h]
\caption{
The $\phi$
distributions for the a) $f_0(1370)$,
b) $f_0(1500)$, c) $f_2(1270)$ and d) $f_2(1950)$
for the data (dots) and the Monte Carlo (histogram).
}
\label{fi:4}
\end{figure}
\begin{center}
\epsfig{figure=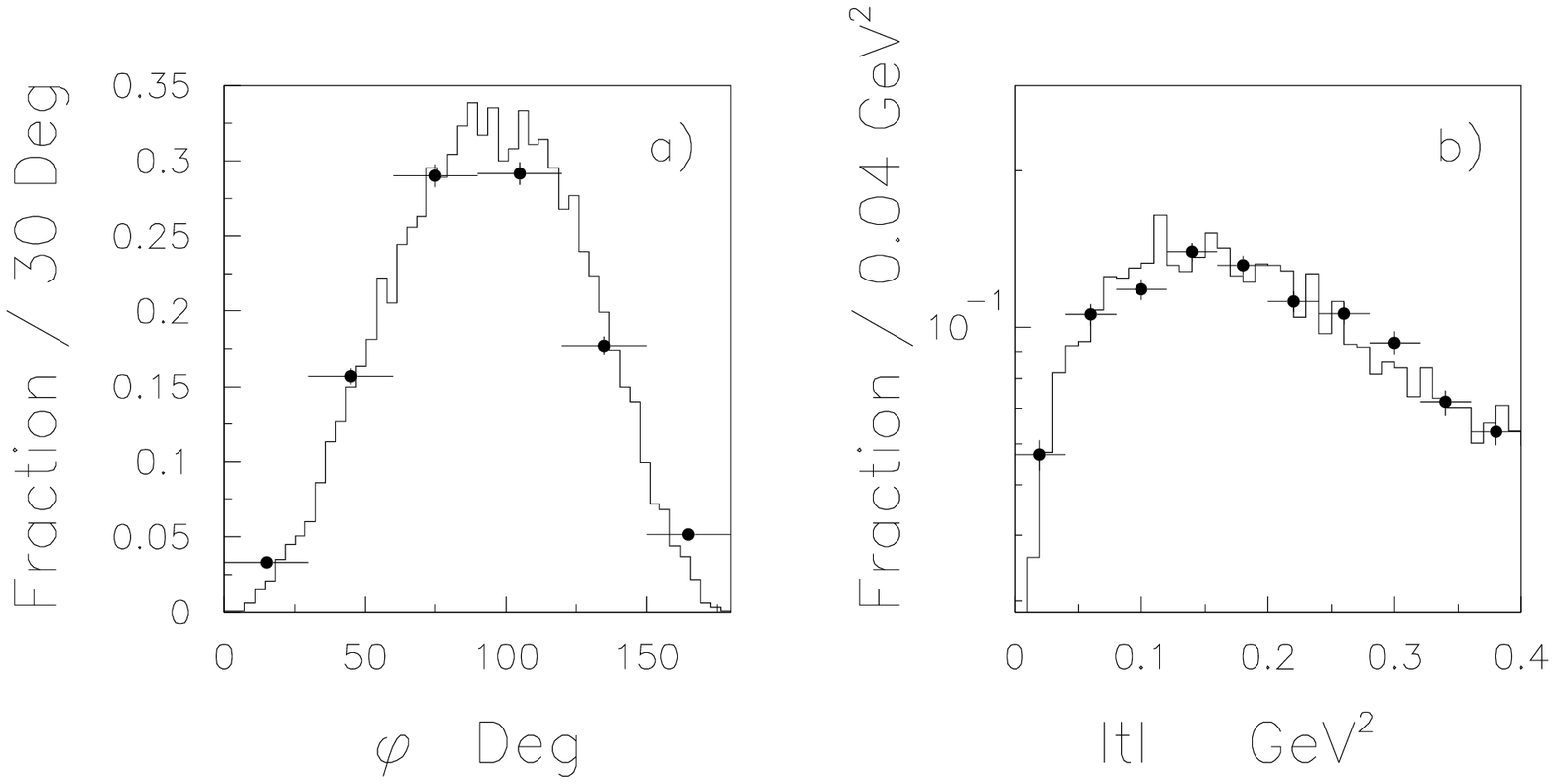,height=22cm,width=17cm}
\end{center}
\begin{center} {Figure 1} \end{center}
\newpage
\begin{center}
\epsfig{figure=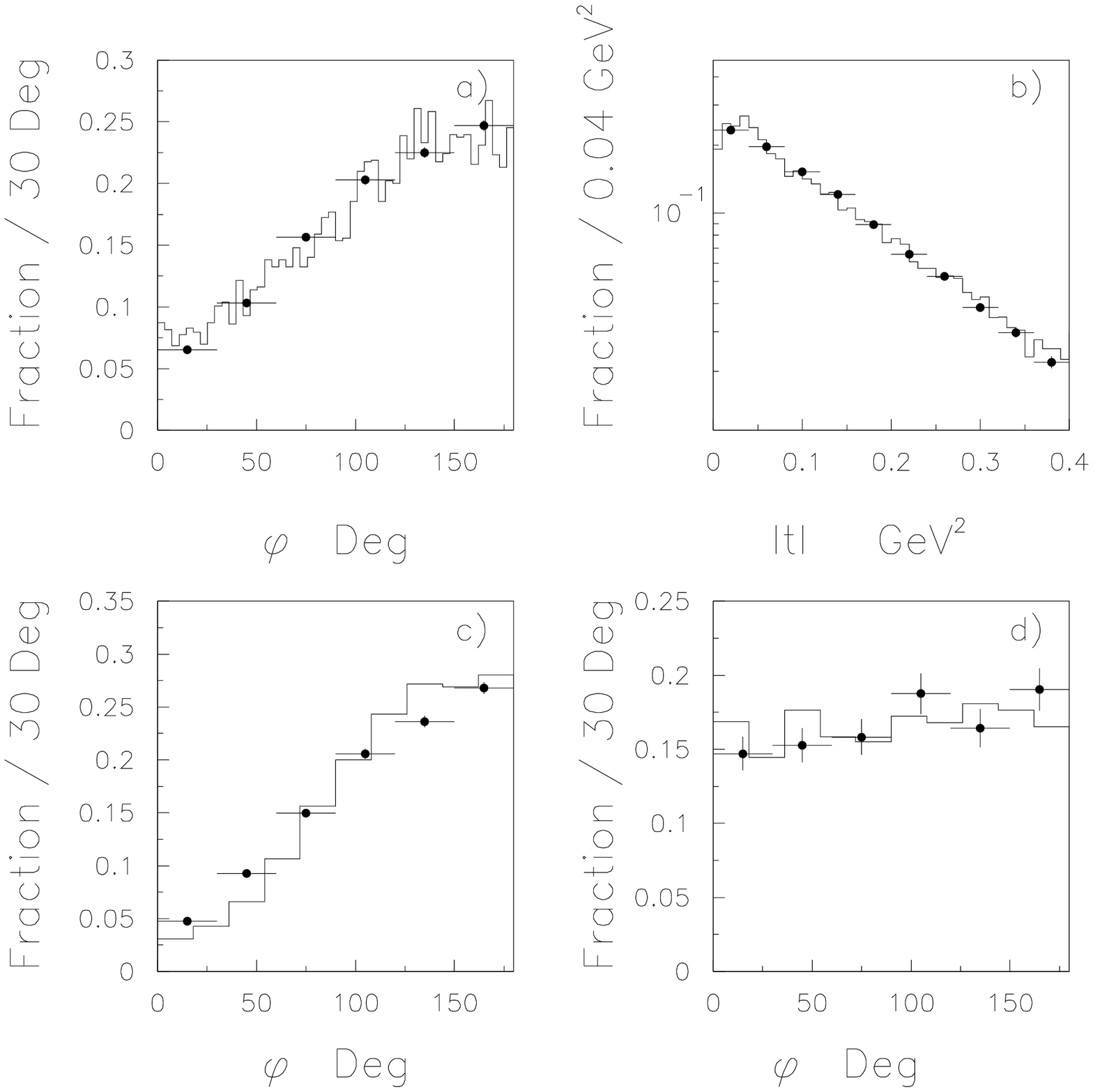,height=22cm,width=17cm}
\end{center}
\begin{center} {Figure 2} \end{center}
\newpage
\begin{center}
\epsfig{figure=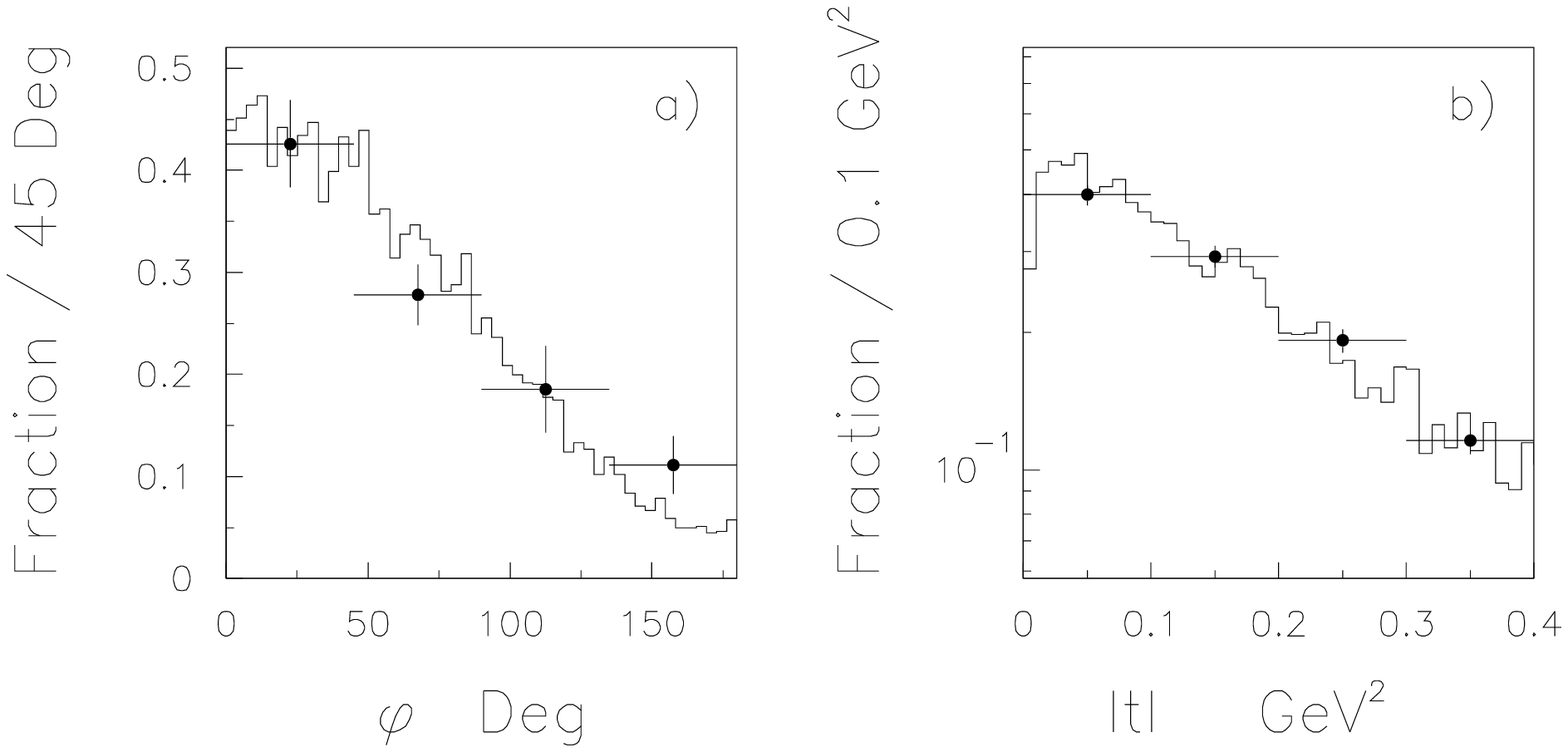,height=22cm,width=17cm}
\end{center}
\begin{center} {Figure 3} \end{center}
\newpage
\begin{center}
\epsfig{figure=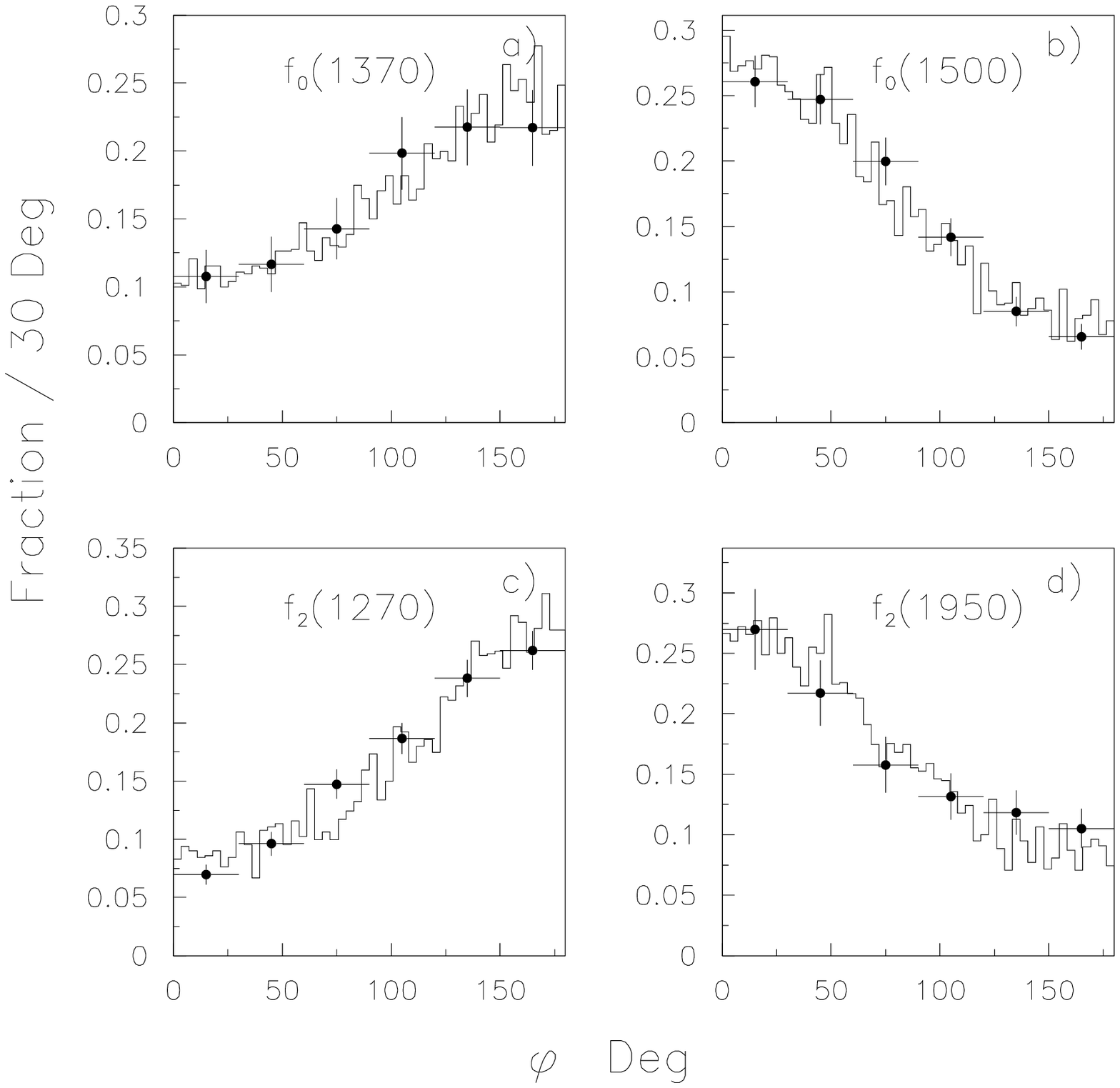,height=22cm,width=17cm}
\end{center}
\begin{center} {Figure 4} \end{center}
\end{document}